\documentclass[prd,amsmath,superscriptaddress,twocolumn]{revtex4}

\usepackage{hyperref}
\usepackage{ifthen}
\usepackage{amsmath}
\usepackage{amssymb}
\usepackage{graphics}
\usepackage{color,bm}

\definecolor{darkgreen}{cmyk}{0.85,0.2,1.00,0.2}

\newcommand{\bq}{\begin{equation}}
\newcommand{\eq}{\end{equation}}
\newcommand{\bqa}{\begin{eqnarray}}
\newcommand{\eqa}{\end{eqnarray}}

\newcommand{\Om}{\Omega_{\rm m}}
\newcommand{\Olam}{\Omega_{\Lambda}}

\newcommand{\Msunh}{M_\odot /h}
\newcommand{\hMpc}{h^{-1}~\text{Mpc}}
\newcommand{\absfR}{|f_{R0}|}
\newcommand{\rmd}{\ensuremath{\mathrm{d}}}
\newcommand{\rhom}{\rho_{\rm m}}
\newcommand{\brhom}{\bar{\rho}_{\rm m}}
\newcommand{\drhom}{\delta\rhom}
\newcommand{\Mvir}{M_{\rm vir}}
\newcommand{\rhoc}{\bar{\rho}_{c}}
\newcommand{\dfR}{\delta f_R}
\newcommand{\RTH}{R_{\rm TH}}
\newcommand{\rhoin}{\rho_{\rm in}}
\newcommand{\rhoout}{\rho_{\rm out}}
\newcommand{\yhal}{y_{\rm h}}
\newcommand{\yenv}{y_{\rm env}}
\newcommand{\deltac}{\delta_{\rm c}}

\newcommand{\deltacLCDM}{\delta_{\rm c}^{\Lambda}}
\newcommand{\denv}{\delta_{\rm env}}
\newcommand{\zform}{z_{\rm c}}

\begin{document}

\title{Modeling halo mass functions in chameleon $f(R)$ gravity}

\author{Lucas~Lombriser}
\affiliation{Institute of Cosmology \& Gravitation, University of Portsmouth, Portsmouth, PO1 3FX, UK}
\author{Baojiu~Li}
\affiliation{Institute for Computational Cosmology, Physics Department, University of Durham, Durham, DH1 3LE, UK}
\author{Kazuya~Koyama}
\affiliation{Institute of Cosmology \& Gravitation, University of Portsmouth, Portsmouth, PO1 3FX, UK}
\author{Gong-Bo~Zhao}
\affiliation{National Astronomy Observatories, Chinese Academy of Science, Beijing, 100012, P.~R.~China}
\affiliation{Institute of Cosmology \& Gravitation, University of Portsmouth, Portsmouth, PO1 3FX, UK}

\date{\today}

\begin{abstract}
On cosmological scales, observations of the cluster abundance currently place the strongest constraints on $f(R)$ gravity.
These constraints lie in the large-field limit, where the modifications of general relativity can correctly be modeled by setting the Compton wavelength of the scalar field to its background value.
These bounds are, however, at the verge of penetrating into a regime where the modifications become nonlinearly suppressed due to the chameleon mechanism and cannot be described by this linearized approximation.
For future constraints based on observations subjected to cluster abundance, it is therefore essential to consistently model the chameleon effect.
We analyze descriptions of the halo mass function in chameleon $f(R)$ gravity using a mass- and environment-dependent spherical collapse model in combination with excursion set theory and phenomenological fits to $N$-body simulations in the $\Lambda$CDM and $f(R)$ gravity scenarios.
Our halo mass functions consistently incorporate the chameleon suppression and cosmological parameter dependencies, improving upon previous formalisms and providing an important extension to $N$-body simulations for the application in consistent tests of gravity with observables sensitive to the abundance of clusters.
\end{abstract}

\maketitle


\section{Introduction}

In $f(R)$ gravity, the Einstein-Hilbert action is supplemented with a free nonlinear function $f(R)$ of the Ricci scalar $R$~\cite{buchdahl:70}, which when designed appropriately can contribute to produce the observed late-time accelerated expansion of our Universe~\cite{carroll:03, nojiri:03, capozziello:03}.
$f(R)$ gravity is formally equivalent to a scalar-tensor theory where the additional degree of freedom is described by the \emph{scalaron} field $f_R \equiv \rmd f/\rmd R$~\cite{starobinsky:79, starobinsky:80, maeda:87} and the kinetic coupling vanishes in Jordan frame.
Here, we parametrize our models by the scalaron field evaluated at the present background, $\absfR$.
The $f_R$ field is massive, and below its Compton wavelength, it enhances gravitational forces by a factor of $1/3$, increasing the growth of structure.
Due to the density dependence of the scalaron's mass, $f(R)$ gravity models may incorporate the \emph{chameleon} suppression~\cite{khoury:03, navarro:06, faulkner:06}, returning gravitational forces to Newtonian relations in high-density regions and making them compatible with Solar System tests~\cite{hu:07a, brax:08}.

The enhanced gravitational coupling at low curvature and below the Compton wavelength can be utilized to place constraints on the $f(R)$ modification.
The transition required to interpolate between the low curvature of the large-scale structure and the high curvature of the galactic halo~\cite{hu:07a} as well as the comparison of nearby distances inferred from Cepheids and tip of the red giant branch stars in a sample of unscreened dwarf galaxies~\cite{jain:12} set the currently strongest bounds on the background field, $\absfR<|\Psi|\sim (10^{-7} - 10^{-5})$, i.e., the typical depth of cosmological potential wells.
Independently, strong constraints can also be inferred from the cosmological structure only.
In this large-scale regime, the currently strongest constraints on $f(R)$ gravity models are inferred from the analysis of the abundance of clusters, yielding a constraint of $\absfR\lesssim10^{-4}$~\cite{schmidt:09, lombriser:10, ferraro:10, lombriser:11b}.

It is important to note that the cluster-scale constraints have been derived by relying on a \emph{linearized} approach of the $f(R)$ modifications, assuming a linear relation between the curvature fluctuation $\delta R$ and the field fluctuation $\delta f_R$ that is correctly described by the background Compton wavelength of the scalaron.
This approach breaks down when $\absfR \lesssim 10^{-5}$, where cluster scales are affected by the chameleon suppression.
It is therefore important for comparison to future measurements to describe the observable quantities encompassing the chameleon effect.
While $N$-body simulations provide a great laboratory for the study of the chameleon mechanism~\cite{oyaizu:08a, oyaizu:08b, zhao:10b, li:11, li:12c, jennings:12, puchwein:13}, semianalytic models need to be developed based on these simulations, in order to allow for a full exploration of the cosmological parameter space in the model comparison to observations~\cite{li:11b, lombriser:12, borisov:11, li:11a, li:12b, lam:12}.

In this paper, we develop and compare different prescriptions for modeling the halo mass function in chameleon $f(R)$ gravity based on the mass- and environment-dependent spherical collapse for chameleon theories~\cite{li:11a} applied to $f(R)$ models in combination with excursion set theory and phenomenological fits to $N$-body simulations.
Our descriptions of the halo mass function incorporate the chameleon mechanism and cosmological parameter dependencies and show good agreement with $N$-body simulations and previous fitting formulae for $f(R)$ gravity without the need of introducing new fitting parameters for the chameleon modification.
Thus, they are well suited for complementing $N$-body simulations for the consistent comparison of $f(R)$ gravity to observations that are sensitive to the abundance of clusters.

The outline of the paper is as follows.
In~\textsection\ref{sec:fRgravity}, we review $f(R)$ gravity with a particular focus on the Hu-Sawicki model~\cite{hu:07a}.
We discuss the linearized and suppressed regimes and a description of the transition between them by an estimation of the thin-shell thickness~\cite{khoury:03}.
In~\textsection\ref{sec:structure}, we examine the evolution and formation of structure in $f(R)$ gravity, in specific, through the spherical collapse model for chameleon theories and extended excursion set theory with a conditional moving barrier.
We further give here details about the $N$-body simulations employed.
\textsection\ref{sec:clusterproperties} is devoted to the modeling of the halo mass function in chameleon $f(R)$ gravity and the comparison of the different approaches based on the spherical collapse model, excursion set theory, and phenomenological fitting functions to $N$-body simulations.
We conclude in~\textsection\ref{sec:conclusions}.


\section{$f(R)$ gravity} \label{sec:fRgravity}

In $f(R)$ gravity, the Einstein-Hilbert action is supplemented with a free nonlinear function of the Ricci scalar $R$,
\begin{equation}
S = \frac{1}{2\kappa^2} \int \rmd^4 x \sqrt{-g} \left[ R + f(R) \right] + S_{\rm m}\left(\psi_{\rm m}; g_{\mu\nu}\right),
 \label{eq:jordan}
\end{equation}
where, $\kappa^2 \equiv 8\pi \, G$, $S_{\rm m}$ is the matter action with matter fields $\psi_{\rm m}$, and we have adopted natural units.
We specialize here to metric $f(R)$ gravity, where the connection is of Levi-Civita type and the modified Einstein field equations are obtained as usual from the variation of the action Eq.~(\ref{eq:jordan}) with respect to the metric $g_{\mu\nu}$, 
\begin{equation}
G_{\mu\nu} + f_R R_{\mu\nu} - \left( \frac{f}{2} - \Box f_R \right) g_{\mu\nu} - \nabla_{\mu} \nabla_{\nu} f_R = \kappa^2 \, T_{\mu\nu}.
\end{equation}
The scalaron $f_R \equiv \rmd f/\rmd R$ is the additional scalar degree of freedom of the model, characterizing the modification of the gravitational force.

We further concentrate on the functional form of $f(R)$ proposed by Hu \& Sawicki~\cite{hu:07a},
\begin{equation}
f(R) = -\bar{m}^2 \frac{c_1 \left( R/\bar{m}^2 \right)^n}{c_2 \left( R/\bar{m}^2 \right)^n + 1}.
\label{eq:husawicki}
\end{equation}
Here, $\bar{m}^2 \equiv \kappa^2 \, \bar{\rho}_{{\rm m}0} / 3$ and overbars refer to background quantities.
We require the modification $f(R)$ to satisfy Solar System tests~\cite{hu:07a} through the chameleon mechanism~\cite{khoury:03, navarro:06, faulkner:06} and moreover, yield a Hubble parameter that matches the $\Lambda$CDM expansion history.
This constrains the free parameters of the model $c_1$, $c_2$, and $n$.
At high curvatures, $c_2^{1/n} R \gg \bar{m}^2$ and Eq.~(\ref{eq:husawicki}) simplifies to
\begin{equation}
 f(R) \simeq -\frac{c_1}{c_2} {\bar m}^2 - \frac{f_{R0}}{n} \frac{\bar{R}_0^{n+1}}{R^n},
 \label{eq:backgroundmimick}
\end{equation}
with $f_{R0} \equiv f_R(\bar{R}_0)$ and $\bar{R}_0$ denoting the present background curvature.
Furthermore, from requiring $\Lambda$CDM to be recovered when $\absfR \rightarrow 0$, we obtain
\begin{equation}
 \frac{c_1}{c_2} \bar{m}^2 = 2\kappa^2 \, \bar{\rho}_{\Lambda}.
 \label{eq:lambdalimit}
\end{equation}

The scalar field equation follows from the variation of the action Eq.~(\ref{eq:jordan}) with respect to the scalaron and in the quasistatic approximation, for $|f_R|\ll1$, becomes
\bq
 \nabla^2 \delta f_R = \frac{1}{3} \left[ \delta R (f_R) - \kappa^2 \, \delta \rho_{\rm m} \right],
 \label{eq:fR}
\eq
where the background has been subtracted, i.e., $\delta f_R = f_R(R) - f_R(\bar{R})$, $\delta R = R - \bar{R}$, $\delta \rho_{\rm m} = \rho_{\rm m} - \bar{\rho}_{\rm m}$.

The cosmological background is assumed to be spatially homogeneous and isotropic, where we describe the scalar metric perturbations of its Friedmann-Lema\^itre-Robertson-Walker metric by
$\Psi=\delta g_{00}/(2g_{00})$ and $\Phi=\delta g_{ii}/(2g_{ii})$.
The relation of $\Psi$ to the matter density and $\delta R$ is given by the modified Poisson equation~\cite{hu:07a}
\bq
 \nabla^2 \Psi = \frac{2 \kappa^2}{3} \delta\rho_{\rm m} - \frac{1}{6} \delta R (f_R).
 \label{eq:pot}
\eq

\subsection{Linearized and suppressed regimes}

For large values of the background field compared to the typical depth of gravitational potentials, $|f_{R0}|\gg |\Psi| \sim (10^{-7} - 10^{-5})$, we can linearize the field equations, Eqs.~(\ref{eq:fR}) and (\ref{eq:pot}), using the approximation
\begin{equation}
\delta R \approx \left. \frac{\partial R}{\partial f_R} \right|_{R=\bar{R}} \delta f_{R} = 3 m^2 \delta f_R, \label{eq:linapp}
\end{equation}
where $m$ is the mass of the scalaron evaluated at the background and $\lambda_{\rm C} \equiv 2\pi\,m^{-1}$ is its Compton wavelength.
Within this linearized approximation and in Fourier space, the solution to Eqs.~(\ref{eq:fR}) and (\ref{eq:pot}) becomes
\bq
 k^2 \Psi({\bf k}) = -\frac{\kappa^2}{2} \left\{ \frac{4}{3} - \frac{1}{3} \left[ \left( \frac{k}{m\,a} \right)^2 + 1 \right]^{-1} \right\} a^2 \delta \rhom({\bf k}),
 \label{eq:linearized}
\eq
where $k=|{\bf k}|$ is the comoving wavenumber.
From Eq.~(\ref{eq:linearized}), it can be seen that at scales $k \gg m \, a$, gravitational forces are enhanced by a factor of $1/3$.

In the opposite limit, where $|f_{R0}| \ll |\Psi| \sim (10^{-7} - 10^{-5})$, using Eqs.~(\ref{eq:backgroundmimick}), (\ref{eq:lambdalimit}), and (\ref{eq:fR}), in the high-density regions, where $\kappa \, \drhom \gg -3 \nabla^2 \dfR$, the scalar field becomes
\bq
 f_R \simeq f_{R0} \left[ \frac{\bar{R}_0}{\kappa^2(\rhom+4\bar{\rho}_{\Lambda})} \right]^{n+1}.
 \label{eq:interior}
\eq
Hence, for $\rhom \gg \rhoc$ and $n>-1$, we get $f_R \simeq 0$, a suppression of the modifications and a return to Newtonian gravity.
More specifically, in this case, $\delta R = \kappa^2 \drhom$ and the Fourier transform of Eq.~(\ref{eq:pot}) recovers the standard Poisson equation
\bq
 k^2 \Psi({\bf k}) = -\frac{\kappa^2}{2} a^2 \delta \rhom({\bf k}).
 \label{eq:suppressed}
\eq

The transition between the linearized and suppressed regimes, described by Eqs.~(\ref{eq:linearized}) and (\ref{eq:suppressed}), respectively, for a top-hat overdensity may be approximated by an estimation of the thin-shell thickness as we shall discuss in the following.

\subsection{Transition between spherically symmetric shells of constant density} \label{sec:thinshell}

Khoury \& Weltman~\cite{khoury:03} derived an estimation of the radial profile of the scalar field $\varphi$ in a spherically symmetric top-hat overdensity of radius $\RTH$ with constant inner and outer matter density $\rhoin$ and $\rhoout$, respectively.
On the inside and outside of $\RTH$, the solutions of the scalar field, $\varphi_{\rm out}$ and $\varphi_{\rm in}$, minimize the effective scalar field potential $V_{\rm eff}(\varphi)$ defined by the scalar field equation
\bq
 \tilde{\Box}\varphi \equiv V_{\rm eff}'(\varphi)
\eq
with the tilde denoting the Einstein frame.
$V_{\rm eff}$ consists of the scalar field potential $V(\varphi)$ and a contribution from the coupling of $\varphi$ to the matter components.
Ref.~\cite{khoury:03} finds that the distance that is necessary for $\varphi$ to settle from $\varphi_{\rm out}$ to $\varphi_{\rm in}$ is approximately given by
\bq
 \frac{\Delta R}{\RTH} \simeq \frac{\kappa}{6 \beta} \frac{\varphi_{\rm out} - \varphi_{\rm in}}{\Psi_{\rm N}},
\eq
where $\beta$ is defined by the transformation of the Jordan frame metric $g_{\mu\nu}$ to the Einstein frame metric $\tilde{g}_{\mu\nu}$ through
\bq
 \tilde{g}_{\mu\nu} = e^{-2\beta\kappa\varphi} g_{\mu\nu}.
\eq
The Newtonian potential at the surface of the sphere is
\bq
 \Psi_{\rm N} = \frac{\kappa^2}{8\pi} \frac{M}{\RTH} = \frac{\kappa^2}{6} \rhoin R_{\rm TH}^2,
\eq
with mass $M \equiv 4\pi \, \rhoin R_{\rm TH}^3$ and hence, we obtain
\bq
 \frac{\Delta R}{\RTH} \simeq \frac{1}{\beta \, \kappa} \frac{\varphi_{\rm out} - \varphi_{\rm in}}{\rho_{\rm in} R_{\rm TH}^2}.
 \label{eq:DRoR}
\eq

In $f(R)$ gravity, $\beta=-1/\sqrt{6}$ and $\tilde{g}_{\mu\nu} = (1+f_R) g_{\mu\nu}$.
Thus, for $|f_R|\ll1$, Eq.~(\ref{eq:DRoR}) becomes
\bq
 \frac{\Delta R}{\RTH} \simeq \frac{3}{\kappa^2 \rho_{\rm in}} \frac{f_{R,{\rm in}} - f_{R, {\rm out}}}{R_{\rm TH}^2}.
\eq
The inner and outer solutions of the scalaron minimizing $V_{\rm eff}(\varphi)$ are equivalent to Eq.~(\ref{eq:interior}), i.e.,
\bq
 f_{R, {\rm in/out}} \simeq \left[ \frac{1+4 \frac{\Olam}{\Om}}{\tilde{\rho}_{\rm in/out} a^{-3} + 4 \frac{\Olam}{\Om}} \right]^{n+1} f_{R0},
\eq
where $\tilde{\rho}_{\rm in/out} \equiv \rho_{\rm m, in/out}(a=1)/\bar{\rho}_{\rm m}(a=1)$.
Therefore, the transition between spherically symmetric shells of constant density in $f(R)$ gravity can approximately be described by
\bqa
 \frac{\Delta R}{\RTH} & \simeq & \frac{\absfR a^3}{\Om \tilde{\rho}_{\rm in} (H_0 \RTH)^2} \left[ \left( \frac{1 + 4\frac{\Olam}{\Om}}{\tilde{\rho}_{\rm out} a^{-3} + 4\frac{\Olam}{\Om}} \right)^{n+1} \right. \nonumber\\
 & & \left. - \left( \frac{1 + 4\frac{\Olam}{\Om}}{\tilde{\rho}_{\rm in} a^{-3} + 4\frac{\Olam}{\Om}} \right)^{n+1} \right]. \label{eq:thinshell}
\eqa

In the thin-shell regime, for $r\in[R_0,\RTH]$, the scalar field is~\cite{khoury:03, li:11a}
\bq
 \varphi(r) \simeq \varphi_{\rm in} + \frac{\kappa\,\beta}{3} \rho_{\rm in} \left( \frac{r^2}{2} + \frac{R_0^3}{r} - \frac{3}{2} R_0^2 \right),
\eq
where $\Delta R=\RTH-R_0$.
Hence, the force enhancement $F$ due to the extra coupling for a unity test mass at $\RTH$ becomes
\bqa
 F \frac{G\,M}{\RTH^2} & \equiv & \left. \kappa\,\beta \nabla \varphi \right|_{\RTH} \nonumber \\
 & \simeq & 2\beta^2 \frac{G\,M}{\RTH^2} \left[ 1 - \left( \frac{R_0}{\RTH} \right)^3 \right] \nonumber \\
 & = & 2\beta^2 \frac{G\,M}{\RTH^2} \left[ 3\frac{\Delta R}{\RTH} - 3\left(\frac{\Delta R}{\RTH}\right)^2 + \left(\frac{\Delta R}{\RTH}\right)^3 \right]. \nonumber \\
 & & \label{eq:extraforce}
\eqa
Note that as $\RTH \geq R_0$ and $R_0\geq0$, we have $\Delta R/\RTH \in [0,1]$, which implies $F\in\left[0,2\beta^2\right]$ and, in specific, $F\in\left[0,1/3\right]$ for $f(R)$ gravity.
Hence, for a top-hat overdensity, Eq.~(\ref{eq:extraforce}) yields an interpolation between the suppressed regime in Eq.~(\ref{eq:suppressed}) and the 1/3 enhancement of the gravitational force in Eq.~(\ref{eq:linearized}), which is $\mathcal{C}^0$ for $\Delta R/\RTH\rightarrow0$ and $\mathcal{C}^2$ for $\Delta R/\RTH\rightarrow1$.

In previous studies~\cite{li:11a,li:12}, only the first term in Eq.~(\ref{eq:extraforce}) has been considered through the approximation
\bq
 F \simeq 2\beta^2\min\left( 3\frac{\Delta R}{\RTH}, 1 \right). \label{eq:extraforceapprox}
\eq
This slightly underestimates the efficiency of the chameleon suppression.
When studying the structure formation in chameleon $f(R)$ gravity through the implementation of the thin-shell approximation in the spherical collapse model in the following, we shall use the full expression Eq.~(\ref{eq:extraforce}).
However, we also study the case of introducing a constant fudge factor $\alpha$ in Eq.~(\ref{eq:extraforceapprox}),
\bq
 F \simeq 2\beta^2\min\left( 3\alpha\frac{\Delta R}{\RTH}, 1 \right), \label{eq:extraforceafudge}
\eq
to modulate the efficiency of the chameleon suppression and account for corrections of approximations such as sphericity~\cite{jones-smith:11} and a top-hat overdensity~\cite{lombriser:12} to realistic structure formation.
Ref.~\cite{li:12} found that a factor of $\alpha\approx1/2$ yields good agreement with the difference between the lensing and dynamical mass of dark matter halos measured in $N$-body simulations of $f(R)$ gravity.


\section{Structure formation} \label{sec:structure}

In the following, we study the formation and evolution of structure in the cold dark matter scenarios of $\Lambda$CDM and $f(R)$ gravity.
We begin by reviewing the linear growth of structure for $\Lambda$CDM and the quasistatic regime of $f(R)$ gravity in~\textsection\ref{sec:lineargrowth}, where due to Eq.~(\ref{eq:linearized}) for $f(R)$ gravity, the linear growth becomes a function of scale in addition to its time dependence.
In~\textsection\ref{sec:sphcoll}, we describe the spherical collapse model for $f(R)$ gravity, discussing its application to excursion set theory in~\textsection\ref{sec:excursion}.
We examine the role of the environmental density in~\textsection\ref{sec:environment} and give details on the $N$-body simulations employed in our study in~\textsection\ref{sec:simulations}.
Note that we focus on cold dark matter halos formed in a $f(R)$ model constructed as alternative to a cosmological constant.
Galaxy clusters in the context of $f(R)$ gravity in the absence of dark matter have been studied, e.g., in~\cite{capozziello:08}.

\subsection{Linear growth of structure} \label{sec:lineargrowth}

In $\Lambda$CDM, combining the linearly perturbed Einstein field equations with energy-momentum conservation, one obtains the ordinary second-order differential equation for the evolution of the matter overdensity $\Delta_{\rm m}(a,k)$ in total matter gauge
\bq
 \Delta_{\rm m}'' + \left[ 2 - \frac{3}{2}\Om(a) \right] \Delta_{\rm m}' - \frac{3}{2}\Om(a) \Delta_{\rm m} = 0, \label{eq:lineargrowtheq}
\eq
where here and throughout the paper, primes denote derivatives with respect to $\ln a$ and $\Om(a) \equiv H_0^2\Om a^{-3}/H^2$.
We define the linear growth function $D(a)$ as
\bq
 \frac{D(a)}{D(a_{\rm i})} \equiv \frac{\Delta_{\rm m}(a,k)}{\Delta_{\rm m}(a_{\rm i},k)} \label{eq:lineargrowth}
\eq
at an initial scale factor $a_{\rm i} \ll 1$ in the matter-dominated regime and solve for $D(a)$ with the corresponding initial conditions $D(a_{\rm i})=a_{\rm i}$ and $D'(a_{\rm i})=a_{\rm i}$.
Note that in this paper, $D(a)$ shall always refer to the linear growth function assuming a $\Lambda$CDM cosmology.

In $f(R)$ gravity, Eq.~(\ref{eq:lineargrowth}) is altered due to the modification of the Poisson equation, Eq.~(\ref{eq:pot}), where for large $|f_R|$, an additional modification of the relation of the lensing potential $(\Phi-\Psi)$ to the matter density fluctuation contributes through the rescaling of its dependency on the matter density by $(1+f_R)^{-1}$.
These modifications can be included in Eq.~({\ref{eq:lineargrowtheq}}) as
\bq
 \Delta_{\rm m}'' + \left[ 2 - \frac{3}{2}\Om(a) \right] \Delta_{\rm m}' - \frac{3}{2} \frac{1-g(a,k)}{1+f_R} \Om(a)  \Delta_{\rm m} \simeq 0, \label{eq:fRlineargrowtheq}
\eq
where as in Eq.~(\ref{eq:linearized})
\bq
 g \equiv \frac{\Phi+\Psi}{\Phi-\Psi} = -\frac{1}{3} \frac{k^2}{k^2+m^2 a^2},
\eq
and correctly describe the time- and scale-dependent linear growth function $D_{f(R)}(a,k)$ defined as in Eq.~(\ref{eq:lineargrowth}) on quasistatic scales~\cite{lombriser:10}.
Note, however, that at near-horizon scales, for scalar-tensor models like $f(R)$ gravity, the matter fluctuation obtained from combining the linearly perturbed Einstein equations with energy-momentum conservation, in general, deviates from the matter fluctuation inferred from the quasistatic description Eq.~(\ref{eq:fRlineargrowtheq})~\cite{lombriser:13}.
Since this modification is small in $f(R)$ gravity and additionally, here, we are interested in the high-curvature regime, we can safely neglect this contribution and furthermore the lensing modification as we restrict to models where $|f_R|\ll1$.

\subsection{Spherical collapse}  \label{sec:sphcoll}

We study the formation of clusters in $f(R)$ gravity using the spherical collapse model.
We approximate the dark matter halo by a spherically symmetric top-hat overdensity of initial radius $\RTH$ with a constant matter density $\rhoin$ and $\rhoout$ on the inside and outside.
In order to incorporate the chameleon suppression in the spherical collapse calculation, we follow~\cite{li:11a} and implement the thin-shell thickness estimator for the chameleon transition by~\cite{khoury:03} described in~\textsection\ref{sec:thinshell} in the case of $f(R)$ gravity.
We introduce $\xi(a)$ to denote the physical radius of the overdensity at $a$, where $\xi(a_{\rm i})=a_{\rm i}\RTH$.
Note that the nonlinear evolution of this overdensity causes $\xi(a)$ to deviate from this simple linear relation at $a>a_{\rm i}$.
This deviation shall be denoted by the dimensionless variable $y\equiv \xi(a)/a\RTH$.
Conservation of mass enclosed in the overdensity implies $\brhom a^3 R_{\rm TH}^3 = \rhom \xi^3$ and hence, $\tilde{\rho} = \rhom/\brhom = y^{-3}$.

From Eq.~(\ref{eq:thinshell}), it follows that the thickness of the thin shell is
\bqa
 \frac{\Delta \xi}{\xi} & \simeq & \frac{\absfR \, a^{3n+4}}{\Om(H_0\RTH)^2} \yhal \left[ \left( \frac{1+4\frac{\Olam}{\Om}}{\yenv^{-3} + 4\frac{\Olam}{\Om} a^3} \right)^{n+1} \right. \nonumber\\
 & & \left. - \left( \frac{1+4\frac{\Olam}{\Om}}{\yhal^{-3} + 4\frac{\Olam}{\Om} a^3} \right)^{n+1} \right], \label{eq:thinshellsphcoll}
\eqa
where we use the notation $\yhal$ and $\yenv$ to refer to the inner and outer overdensities, the halo and its local environment, respectively.

In order to describe the evolution of $\yhal$, we model the effective modification to Newton's constant as
\bq
 G_{\rm eff} = \left[1 + F\left( \frac{\Delta\xi}{\xi} \right) \right] G,
\eq
where $F(\Delta\xi/\xi)$ is given by the thin-shell approximation in~\textsection\ref{sec:thinshell}.
With this modification, the equation of motion of the spherical shell is given by~\cite{li:11a,schmidt:08}
\bq
 \frac{\ddot{\xi}}{\xi} = -\frac{\kappa^2}{6} \left( \brhom - 2\bar{\rho}_{\Lambda} \right) - \frac{\kappa^2}{6} (1 + F) \drhom,
 \label{eq:eqmot}
\eq
which with $\tilde{\rho}_{\rm in}=y_{\rm h}^{-3}$ yields
\bq
 \yhal'' + \left[ 2 - \frac{3}{2} \Om(a) \right] \yhal' + \frac{1}{2} \Om(a) (1+F) \left( \yhal^{-3} - 1 \right) \yhal = 0, \label{eq:yhal}
\eq
where dots denote cosmic time derivatives.
Note that in Eq.~(\ref{eq:eqmot}), we have used the contribution of the modified force through $F\,\drhom$ rather than through $F\rhom$, which was used in~\cite{li:11a}.
The final expression for the evolution of $\yhal$ in Eq.~(\ref{eq:yhal}), however, is in agreement with Eq.~(35) of~\cite{li:11a}.

We assume that the environment follows a $\Lambda$CDM evolution, which in Eq.~(\ref{eq:eqmot}) is obtained in the limit $\Delta\xi/\xi\rightarrow0$ or equivalently, $F\rightarrow0$.
Thus,
\bq
 \yenv'' + \left[ 2 - \frac{3}{2} \Om(a) \right] \yenv' + \frac{1}{2} \Om(a) \left( \yenv^{-3}-1 \right) \yenv = 0. \label{eq:yenv}
\eq

Eqs.~(\ref{eq:yhal}) and (\ref{eq:yenv}) form a system of coupled differential equations, which we solve by setting the initial conditions at $a_{\rm i} \ll 1$ in the matter-dominated regime,
\bqa
 y_{\rm h/env, i} & = & 1 - \frac{\delta_{\rm h/env, i}}{3}, \\
 y_{\rm h/env, i}' & = & - \frac{\delta_{\rm h/env, i}}{3}.
\eqa
We use the $\Lambda$CDM linear growth function $D(a)$ from Eq.~(\ref{eq:lineargrowth}) to extrapolate initial overdensities to present time, defining an effective linear overdensity
\bq
 \delta_{\rm h/env}({\bf x}; \xi_{\rm h/env}) \equiv \frac{D(1)}{D(a_{\rm i})} \delta_{\rm h/env, i}. \label{eq:extrapolation}
\eq

\subsection{Excursion set theory} \label{sec:excursion}

Excursion sets correspond to regions where the matter density smoothed over this region exceeds a given threshold, defining the regions where virialized structures are expected to have formed~\cite{schaeffer:85, cole:88, cole:89, efstathiou:88a, efstathiou:88b, narayan:87, carlberg:88}.
The smoothed matter density perturbation field over a region of radius $R$ is
\bqa
 \delta(\mathbf{x},R) & = & \int W(|\mathbf{x}-\mathbf{y}|;R) \delta(\mathbf{y}) \rmd^3\mathbf{y} \nonumber\\
  & = & \int \tilde{W}(k;R) \delta_{\mathbf{k}} e^{i \mathbf{k}\cdot\mathbf{x}} \rmd^3\mathbf{k}, \label{eq:smoothdelta}
\eqa
where $W(|\mathbf{x}-\mathbf{y}|;R)$ is a window function and $\delta(\mathbf{x})\equiv \rhom(\mathbf{x})/\brhom -1$ is the matter density perturbation. $\tilde{W}(k;R)$ and $\delta_{\mathbf{k}}$ are the corresponding Fourier transforms.
For now, we shall only consider the initial density perturbation field and use $\delta(\mathbf{x})$ to refer to it.
We assume $\delta(\mathbf{x})$ to be Gaussian.
The density fluctuation field is characterized by its power spectrum $P(k)$, for which the variance is
\bq
 S(R) \equiv \sigma^2(R) \equiv \langle \delta^2(\mathbf{x};R) \rangle = \int \tilde{W}(k;R) P(k) \rmd^3\mathbf{k}. \label{eq:variance}
\eq
Hence, given the power spectrum, one can interchange $S$ and $R$ as measures of the scale of spherical perturbations.
For a sharp window function $\tilde{W}(k;R)$ in $k$-space, an incremental step in the smoothed initial overdensity field $\delta(\mathbf{x};R)$ in Eq.~(\ref{eq:smoothdelta}) is attributed to the extra higher-$k$ modes.
In this case, the wavenumbers are uncorrelated such that the incremental steps satisfy the Markov property, i.e., steps depend on the current value only and are independent of previous values.
The increment is a Gaussian field with zero mean and variance $\rmd S$ such that $\delta(\mathbf{x};S)$ can be described by a Brownian motion in $S$ with Gaussian probability distribution
\bq
 P(\delta,S) \rmd \delta = \frac{1}{\sqrt{2\pi \, S}} e^{-\delta^2/2S} \rmd \delta.
\eq
For a scale-independent linear growth function, determining the growth of both $\delta$ and $\sqrt{S}$, the linear density field remains Gaussian at all times.
This holds particularly for a $\Lambda$CDM universe and from now on, $\delta(\mathbf{x};R)$ shall refer to the extrapolation of the smoothed initial matter density perturbation to $z=0$ via the linear density growth function $D(a)$ from Eq.~(\ref{eq:lineargrowth}) [see Eq.~(\ref{eq:extrapolation})].
Note that since the linear growth function is simpler to calculate for $\Lambda$CDM than for $f(R)$ gravity and especially due to its scale independence, we shall always use the linear $\Lambda$CDM growth function $D(a)$ to do this extrapolation.
Hence, for $f(R)$ gravity, the extrapolated matter density field and associated quantities should be interpreted as effective quantities only.

\begin{figure}
 \resizebox{\hsize}{!}{\includegraphics{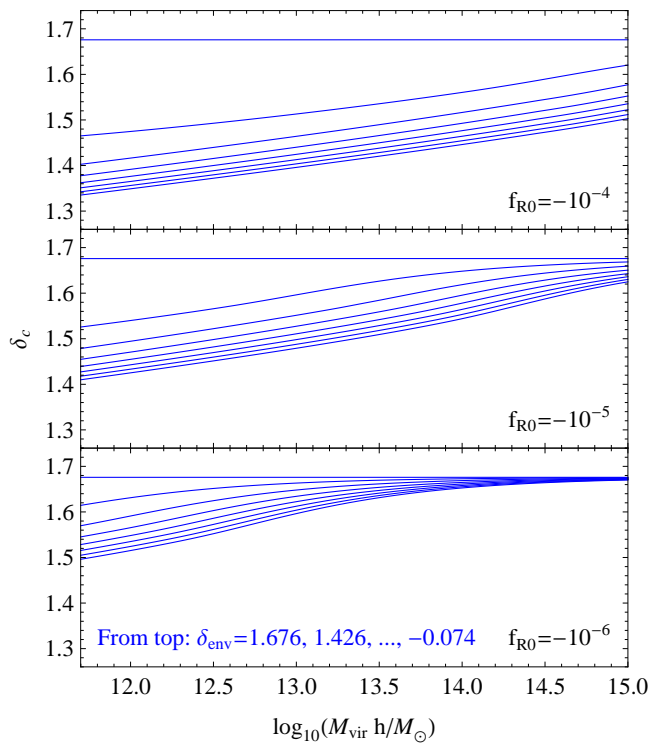}}
 \caption{
The collapse density $\deltac$ in chameleon $f(R)$ gravity predicted by the spherical collapse model for different $\absfR$ at $\zform=0$.
The chameleon effect is incorporated via Eq.~(\ref{eq:extraforce}) and the thin-shell thickness Eq.~(\ref{eq:thinshellsphcoll}).
Note that the predictions for $\deltac$ in $f(R)$ gravity return to the $\Lambda$CDM value $\deltacLCDM\simeq1.676$ when the environmental density fluctuation approaches the value of the halo overdensity.
}
\label{fig:dc}
\end{figure}

In this spirit, a spherical region of initial radius $R$ is considered to have collapsed to a virialized object today or live in a larger region which has collapsed earlier if $\delta(\mathbf{x};\geq R) \geq \delta_{\rm c}$, where $\delta_{\rm c}$ is the effective \emph{collapse density}, which may be determined from the initial matter overdensity causing a singularity in Eq.~(\ref{eq:yhal}) and extrapolated to present time via $D(a)$.
In $f(R)$ gravity, this critical density is dependent on the mass of the top-hat overdensity and the local environment, $\deltac(\mathbf{x};M,\denv)$ with $M\approx 4\pi \bar{\rho}_{\rm m 0} R_{\rm TH}^3$.
We show the mass dependency of $\deltac$ for different $\absfR$ and $\denv$ for collapse today in Fig.~\ref{fig:dc}.
In general, in addition to the top-hat mass and the environment, $\deltac$ is dependent on $\Om$ and the redshift of the collapse $\zform$.

In a $\Lambda$CDM universe, $\deltac$ becomes independent of mass and environment and thus, for a given $\Om$ at a given $\zform$, defines a flat barrier $\deltacLCDM$.
In this case, the fraction of mass enclosed in virialized dark matter halos of mass $M\geq4\pi\,R^3\bar{\rho}_{\rm m, i}$ to the total mass is~\cite{bond:91}
\bq
 \mathcal{F}(M,z) = \frac{1}{\sqrt{2\pi S}} \int_{\frac{D(0)}{D(z)}\deltacLCDM}^{\infty} \left[ e^{-\delta^2/2S} - e^{-\left(\delta-\deltacLCDM\right)^2/2S} \right] \rmd \delta.
\eq
This corresponds to the fraction of Brownian motion trajectories which have crossed $\deltacLCDM$ at $S$.
The fraction of mass enclosed in halos of masses corresponding to $S(M)\in[S,S+\rmd S]$ that collapse at $z=\zform$ is given by the Press-Schechter expression~\cite{press:74}
\bq
 \phi(S,\zform) \rmd S = \frac{1}{\sqrt{2\pi\,S}} \frac{D(0)}{D(\zform)} \frac{\deltacLCDM}{S} \exp \left[ - \frac{1}{2} \frac{D(0)}{D(\zform)} \frac{\left(\deltacLCDM\right)^2}{S} \right] \rmd S, \label{eq:fs}
\eq
where $\phi(S)$ describes the distribution of Brownian motion trajectories that first cross the barrier $D(z=0) \deltacLCDM / D(z=\zform)$ at $S$.

In $f(R)$ gravity the barrier $\deltac$ is no longer flat and becomes dependent on $S$ and the environment embedding the collapsing halo.
In order to characterize the environment, we follow~\cite{li:11a} and study a top-hat sphere with density perturbation $\delta_{\rm env}(\mathbf{x};\chi)$, evolving according to $\Lambda$CDM and specified by the choice of radius $\chi$, which embeds $\delta(\mathbf{x};R)$.
The crossing probability conditional on the Brownian motion trajectory passing $\denv$ at $S_{\chi}$ is $\phi[\deltac(S,\denv),S | \denv,S_{\chi}]$.
This probability needs to be computed numerically, for which we use a code developed in~\cite{li:11a} based on the algorithm of~\cite{parfrey:11}.
We refer to~\cite{li:11a} for more details on this computation.
In the following, we shall clarify the characterization of the environment. 

\subsection{Environment} \label{sec:environment}

The effect of the environment on the first-crossing distribution depends on the definition of the radius $\chi$.
This can be done, for instance, following the \emph{fixed-scale environment approximation} of~\cite{li:11a}, fixing a \emph{Lagrangian} (or initial comoving) radius $\xi$, defining $\chi\equiv\xi$.
We shall adopt the value for the Lagrangian radius used in~\cite{li:11a}, $\xi=8h^{-1}~{\rm Mpc}$ such that $S_{\xi}=\sigma_8^2$.
For comparison to $N$-body simulations or observations that do not differentiate between structures formed in different environments, the conditional first-crossing distribution of the moving barrier $\phi[\deltac(S,\denv),S | \denv,S_{\xi}]$ computed with the algorithm of~\cite{parfrey:11} needs to be integrated over all environments.
In order to do so, in the following, we shall denote the distribution of $\denv$ characterized through $\xi$ as $P_{\xi}(\denv)$, corresponding to the probability that the Brownian motion trajectory passes through $\denv$ at $S_{\xi}$ never having crossed the collapse density $\deltacLCDM$ at $S<S_{\xi}$.
This is given by~\cite{bond:91}
\bqa
 P_{\xi}(\denv) & = & \frac{1}{\sqrt{2\pi\,S_{\xi}}} \Theta(\deltacLCDM-\denv)  \nonumber\\
 & & \times \left[ e^{-\frac{\denv^2}{2S_{\xi}}} - e^{-\frac{(\denv-2\deltacLCDM)^2}{2S_{\xi}}} \right],
\eqa
where $\Theta$ is the Heaviside step function.
The environment-averaged first-crossing distribution becomes
\bq
 \langle \phi(S) \rangle_{\rm env} = \int_{-\infty}^{\deltacLCDM} P_{\xi} \cdot \phi[\deltac(S,\denv),S | \denv,S_{\xi}] \rmd \denv. \label{eq:envavg}
\eq

Note that this reduces to the unconditional first-crossing of a constant barrier $\deltacLCDM$ at $S$,
\bq
 \langle \phi(S) \rangle_{\rm env} = \frac{1}{\sqrt{2\pi\,S}} \frac{\deltacLCDM}{S} e^{-(\deltacLCDM)^2/2S},
\eq
when $\deltac(S,\denv)=\deltacLCDM$, for which
\bq
 \phi[\deltac(S,\denv),S | \denv,S_{\xi}] = \frac{\deltacLCDM-\denv}{\sqrt{2\pi} (S-S_{\xi})^{3/2}} e^{-\frac{\left(\deltacLCDM-\denv\right)^2}{2(S-S_{\xi})}}.
\eq
However, in general, $\langle \phi(S) \rangle_{\rm env}$ must be computed numerically.

A more accurate approach for defining the radius $\chi$ is taken in~\cite{li:12b}, where the size of environments is defined by the Eulerian (physical) radius $\zeta$.
We shall adopt the value used in~\cite{li:12b}, $\zeta=5h^{-1}~{\rm Mpc}$.
We refer to~\cite{li:12b} and in particular Fig.~2 therein for a comparison of the Lagrangian and Eulerian definitions for the environment and implications for the structures inferred from that for chameleon theories.
The Eulerian overdensity at time $t$ is~\cite{bernardeau:94,sheth:98}
\bq
 \Delta_{\rm NL}(t) \simeq \left[ 1 - \frac{\delta(t)}{\deltacLCDM} \right]^{-\deltacLCDM}.
\eq
This defines the barrier for the environment
\bq
 \delta_{\rm env}^{\rm E}(M_{\rm env}) = \deltacLCDM \left[ 1 - \left( \frac{M_{\rm env}}{\brhom \, V} \right)^{-1/\deltacLCDM} \right],
\eq
for which the first crossing thereof determines the $\delta(t)$ that a spherical region containing mass $M_{\rm env}$ must have in order to evolve into an Eulerian volume $V$ at $t$.
For a power-law matter power spectrum $P(k)$ with index $n_s$, this becomes
\bq
 \delta_{\rm env}^{\rm E}(S_{\xi}) = \deltacLCDM  \left[ 1 - \left( \frac{\zeta}{8h^{-1}{\rm Mpc}} \right)^{3/\deltacLCDM} \left( \frac{S_{\xi}}{\sigma_8} \right)^{3/(3+n_s)\deltacLCDM} \right],
\eq
where
\bq
 S_{\xi}(M_{\rm env}) = S(\xi) = \frac{1}{2\pi^2} \int_0^{\infty} k^2 P(k) W^2(k\,\xi) \rmd k
\eq
with Lagrangian radius $\xi$ such that $M_{\rm env}=4\pi\xi^3\brhom/3$.
The first-crossing probability of the moving barrier $\delta_{\rm env}^{\rm E}(S_{\xi})$ in $[S_{\xi},S_{\xi}+\rmd S_{\xi}]$, $P_{\rm env}(S_{\xi}) \rmd S_{\xi}$, corresponds to the probability that an arbitrary point is located in an environment, which will have an Eulerian radius $\zeta$ at $\zform$ and for which $\delta(t) \in [\delta_{\rm env}^{\rm E}(S_{\xi}),\delta_{\rm env}^{\rm E}(S_{\xi}+\rmd S_{\xi})]$.

We use an approximation of the probability distribution of the Eulerian environment $\denv$ by~\cite{lam:08} and also used in~\cite{li:12b, lam:12},
\bqa
 P_{\zeta}(\denv) & = & \frac{\beta^{\omega/2}}{\sqrt{2\pi}} \left[ 1 + (\omega - 1)\frac{\denv}{\deltac} \right] \left( 1 - \frac{\denv}{\deltac} \right)^{-\omega/2-1} \nonumber\\
 & & \times \exp \left[ -\frac{\beta^{\omega}}{2} \frac{\denv}{(1 - \denv/\deltac)^{\omega}} \right], \label{eq:eulerian}
\eqa
where $\beta = (\zeta/8)^3/\deltac / \sigma_8^{2/\omega}$, $\omega = \deltac \gamma$ with
\bq
 \gamma = - \frac{\rmd \ln S_{\xi}}{\rmd \ln M_{\rm env}} = \frac{n_s+3}{3}.
\eq

\subsection{$N$-body simulations} \label{sec:simulations}

Dark matter $N$-body simulations of $f(R)$ gravity provide a great laboratory for studying the chameleon mechanism.
Here, we use simulation results of~\cite{zhao:10b, li:11} for the comparison to the semianalytic modeling described in~\textsection\ref{sec:sphcoll} through \textsection\ref{sec:environment}.
These simulations are performed using a particle mesh code solving the quasistatic relations Eqs.~(\ref{eq:fR}) and (\ref{eq:pot}).
They cover the Newtonian and chameleon scenarios for each field strength $\absfR=10^{-6}, 10^{-5}, 10^{-4}$ with $n=1$ and cosmological parameter values fixed to match WMAP 3-year results, $\Olam=0.76$, $\Om=1-\Olam$, $h=0.73$, $n_{\rm s}=0.958$, and the initial power in curvature fluctuations $A_{\rm s}=(4.89\times 10^{-5})^2$ at $k=0.05~\textrm{Mpc}^{-1}$.
Each set of simulations consists of 10 realizations with each box size, $L_{\rm box} = 64\hMpc, \ 128\hMpc, \ 256\hMpc$, and a total particle number of $N_{\rm p}=256^3$ placed on $128^3$ domain grids.
During the simulation, the domain grids are progressively refined in regions where the local densities are sufficiently large to reach a predefined threshold.
This causes the grid structure to efficiently follow the density distribution so that the high-density regions can be resolved better.
For the identification of halos within the simulation and their associated masses, a spherical overdensity (SO) algorithm (cf.~\cite{jenkins:00}) is used.
Hereby, the particles are placed on the grid by a cloud-in-cell interpolation and counted within a growing sphere around the center of mass until the required overdensity is reached.
The particle masses contained in the sphere then define the mass of the halo.
This process is started at the highest overdensity grid point and hierarchically continued to lower overdensity grid points until all halos are identified.

Note that we use the virial overdensity $\Delta_{\rm vir}$ obtained for $\Lambda$CDM to identify halos even in $f(R)$ gravity in order to make a fair comparison between the different models.
We estimate the error of using $\Lambda$CDM virial masses $\Mvir$ instead of virial masses for $f(R)$ gravity in our predictions with the approximate relation
\bq
 \frac{M_{{\rm vir}, f(R)}}{\Mvir} \simeq \left( \frac{\Delta_{{\rm vir}, f(R)}}{\Delta_{\rm vir}} \right)^{-1/3}, \label{eq:masscorrection}
\eq
which becomes exact for a halo density scaling as $\rho \sim r^{-9/4}$.
This radial dependence for $\rho$ can be motivated for the self-similar secondary infall and accretion in both $\Lambda$CDM and $f(R)$ gravity~\cite{bertschinger:85,lombriser:12}.
For our choice of cosmological parameters, in the case $F=0$, we have $\Delta_{\rm vir}=390$ and for $F=1/3$, the virial overdensity becomes $\Delta_{{\rm vir}, f(R)}=309$~\cite{schmidt:08}.
Hence, the error is approximately 8\% in case of the full modification $F=1/3$.
As this is a very simplified estimate for the mass ratio in Eq.~(\ref{eq:masscorrection}), the full computation requiring mass and environmental dependence of $F$ and the exact radial halo profile, and due to its relative smallness compared to the overall modification, we chose to ignore this effect when comparing models of the halo mass function to the $N$-body results in the following.

Finally, note that recently, it has been shown~\cite{llinares:13} that for symmetron models~\cite{hinterbichler:10}, differences may appear between the scalar field distributions produced in $N$-body simulations when assuming the quasistatic limit and when accounting for time derivatives of the scalar field, respectively.
In the $f(R)$ gravity simulations used here, the scalar field sits at the bottom of the effective potential and never changes sign.
This is different from the symmetron model considered in~\cite{llinares:13}, where the sign of the scalar field can be different in different regions after the symmetry breaking.
Hence, we do not expect the same magnitude in the deviations of the simulation results for $f(R)$ gravity models and we assume that the small-scale structure is correctly described by the quasistatic approximation in Eqs.~(\ref{eq:fR}) and (\ref{eq:pot}).
Furthermore, for $f(R)$ gravity, numerical self-consistency checks have been conducted~\cite{oyaizu:08a}, supporting the assumption of the smallness of the time derivatives.
A more rigorous analysis of the applicability of the quasistatic approximation remains to be conducted in future work.
We refer to~\cite{capozziello:11} for a discussion of time-dependent spherically symmetric perturbations in the Minkowskian limit of $f(R)$ gravity.


\section{Modeling the halo mass function} \label{sec:clusterproperties}

Effects from $f(R)$ modifications of gravity on halo properties have been studied in, e.g.,~\cite{martino:08, schmidt:08, schmidt:10, borisov:11, li:12, li:11b, lombriser:12, lombriser:11b}.
The enhanced abundance of clusters caused by the modification was used in~\cite{schmidt:09, lombriser:10, lombriser:11b} in comparison to observations to place an upper bound on the scalaron background value of $\absfR\lesssim10^{-4}$.
However, given the expected constraints, these analyses have been carried out in the linearized regime of $f(R)$ gravity, where the approximation Eq.~(\ref{eq:linapp}) is valid.
With future measurements, constraints will penetrate into the chameleon regime and it becomes important to consistently incorporate the chameleon effect on the observables.

Here, we focus on describing the halo mass function in $f(R)$ gravity.
Thereby, we use the spherical collapse model, excursion set formalism, and fitting formulae that have been calibrated to $\Lambda$CDM and $f(R)$ gravity $N$-body simulations.
We restrict to $f(R)$ models with exponent $n=1$ corresponding to the choice of $n$ in the $N$-body simulations described in~\textsection\ref{sec:simulations}.
Our relations can be used to explore halo mass functions in the cosmological parameter space beyond the parameter values used in the $N$-body simulations and hence can be applied to consistently constrain $\absfR$ in the chameleon regime.

\begin{figure*}[t!]
 \resizebox{0.495\hsize}{!}{\includegraphics{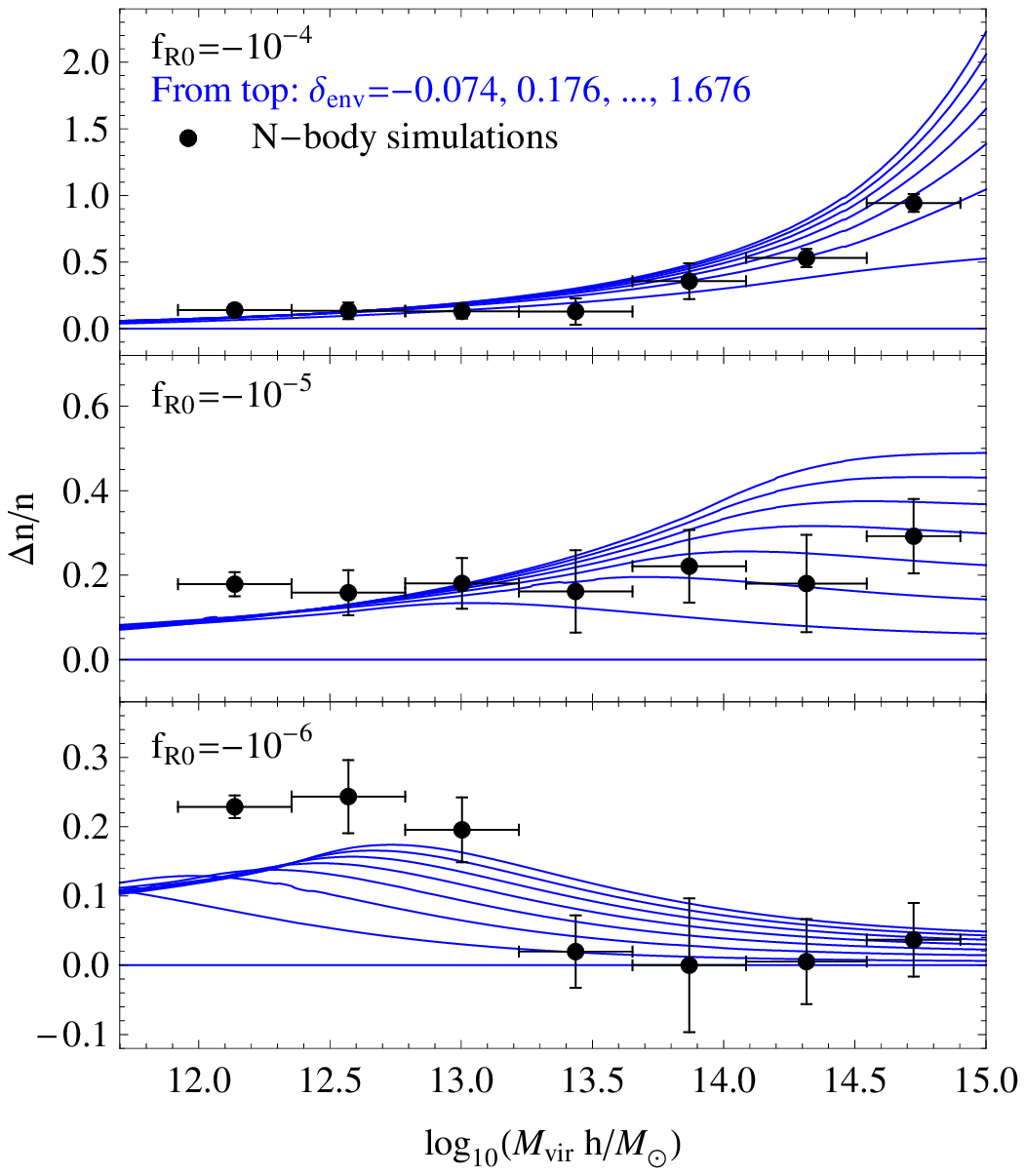}}
 \resizebox{0.495\hsize}{!}{\includegraphics{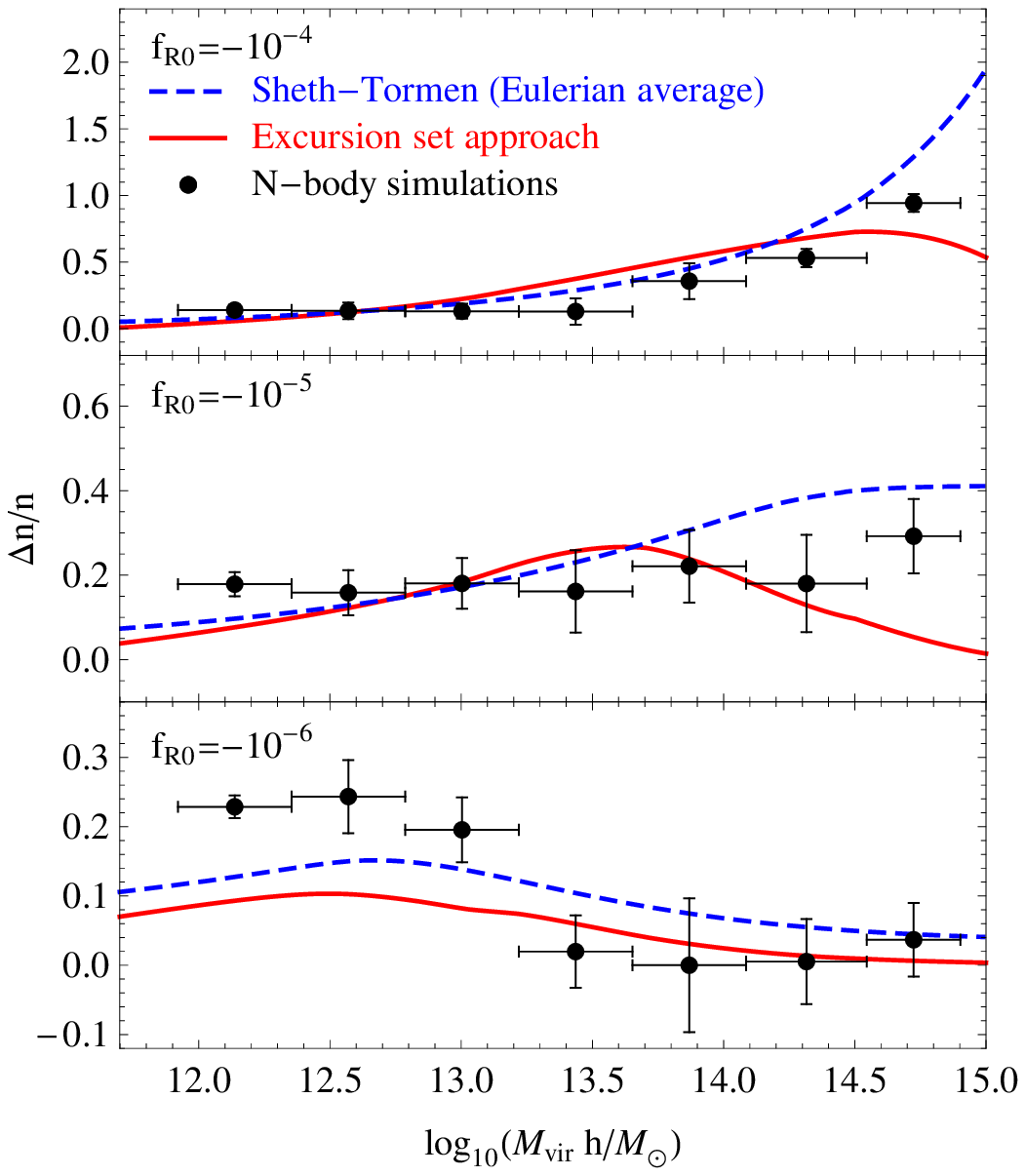}}
\caption{
Relative enhancement of the halo mass function in chameleon $f(R)$ gravity with respect to the prediction for $\Lambda$CDM.
The environmental dependence is illustrated using the collapse density $\deltac$ from Fig.~\ref{fig:dc} computed with the spherical collapse model in~\textsection\ref{sec:sphcoll} and applied to the Sheth-Tormen fit for $\Lambda$CDM simulations Eq.~(\ref{eq:st}) (\emph{left-hand panel}).
These predictions are averaged over the Eulerian environment defined in~\textsection\ref{sec:environment} (dashed line) and compared to the excursion set prediction (solid line) (\emph{right-hand panel}).
Note that the $N$-body results at the low-mass end are contaminated by the inclusion of subhalos, which are not identified and removed in the SO halo-finder employed.
}
\label{fig:massfunctionSTenvalpha}
\end{figure*}

\begin{figure*}[t!]
 \resizebox{0.483\hsize}{!}{\includegraphics{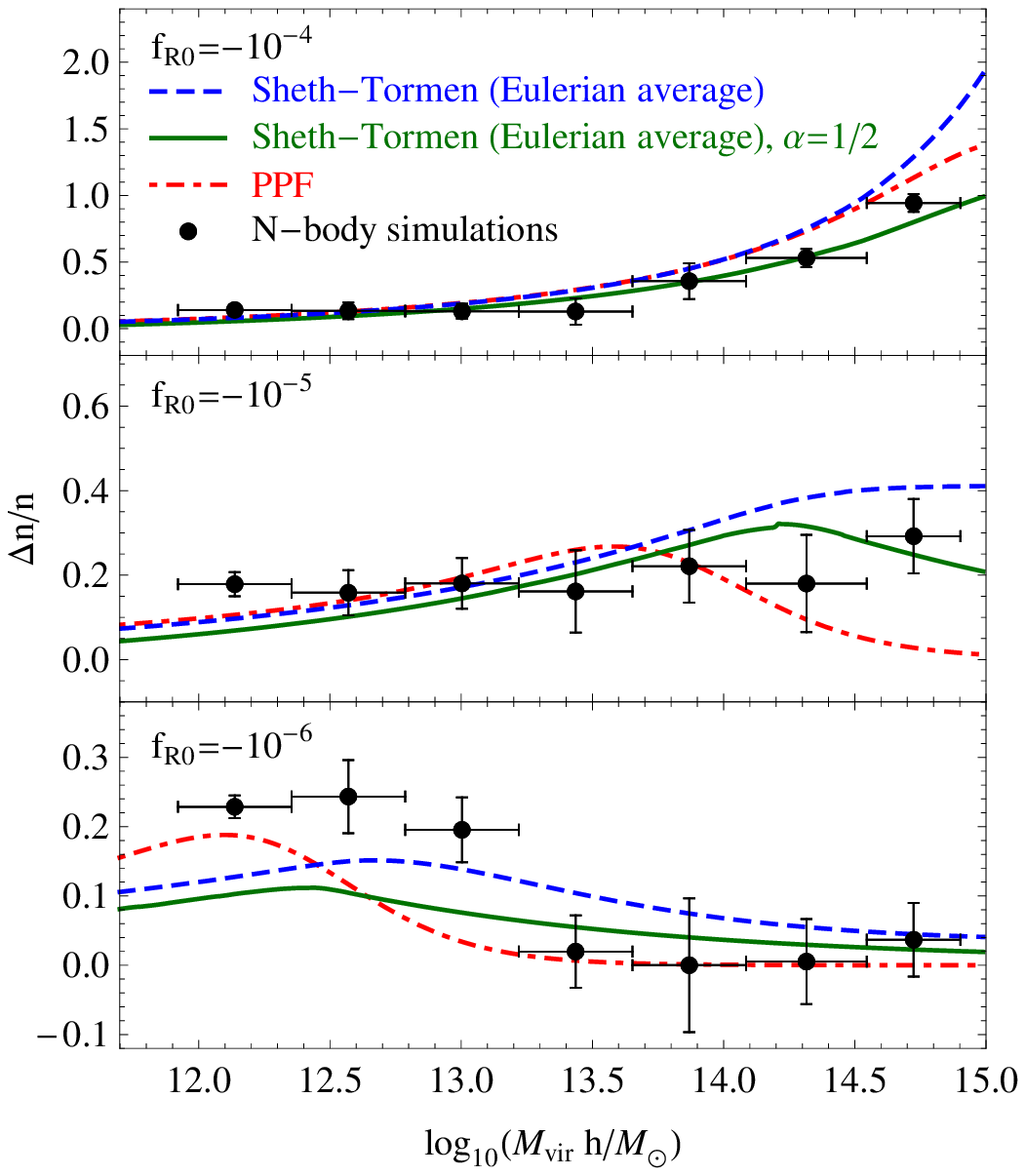}}
 \resizebox{0.507\hsize}{!}{\includegraphics{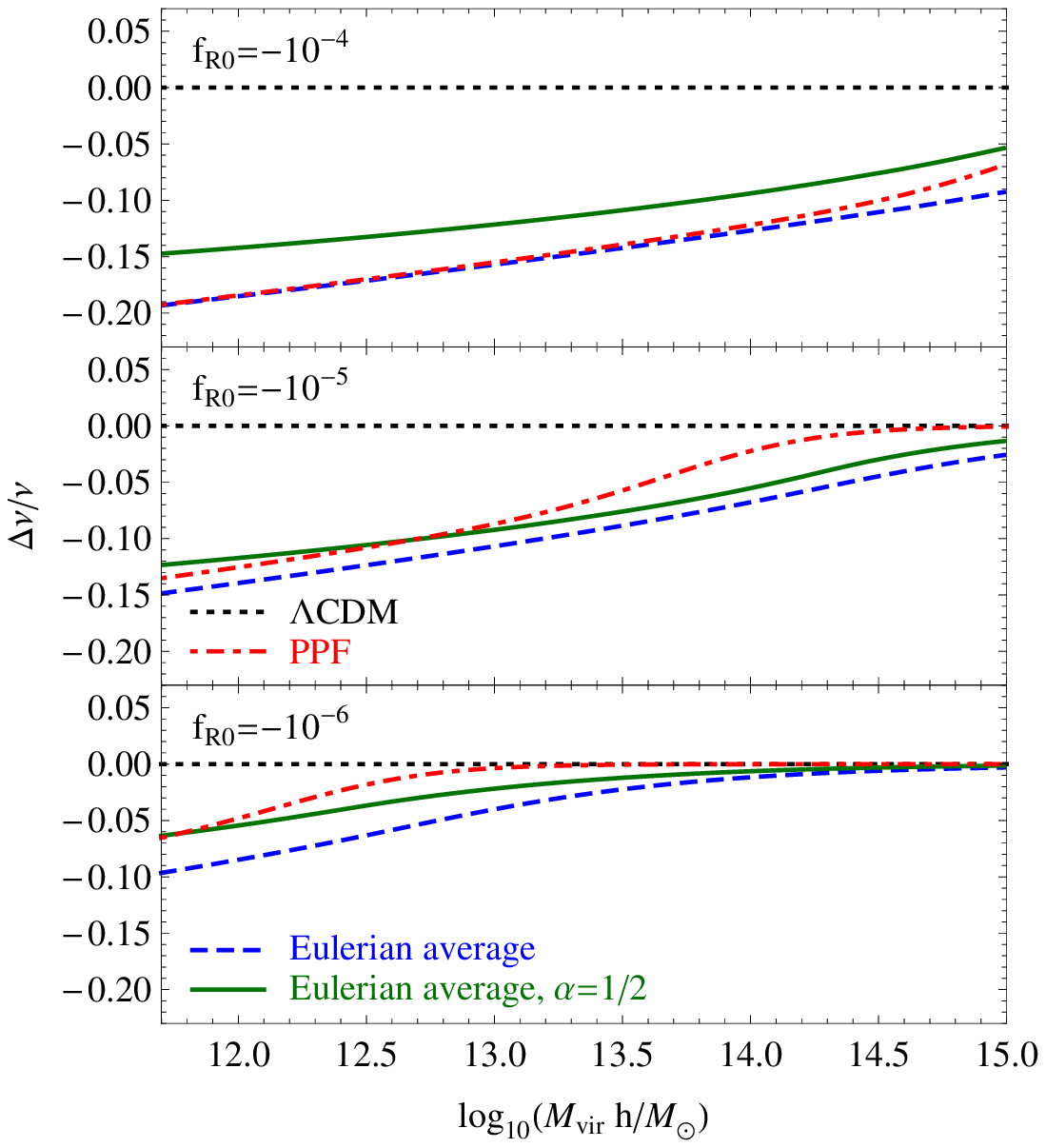}}
\caption{
Comparison of predictions for the relative deviation in the halo mass function (\emph{left-hand panel}) and in the peak threshold (\emph{right-hand panel}) for chameleon $f(R)$ gravity with respect to the $\Lambda$CDM prediction, derived with the spherical collapse model in~\textsection\ref{sec:sphcoll}.
Results are averaged over the Eulerian environment described in~\textsection\ref{sec:environment}, where the halo mass functions are computed using the Sheth-Tormen expression Eq.~(\ref{eq:st}).
The agreement with the $N$-body simulations at the high-mass end can be improved by introducing the fudge factor $\alpha=1/2$ in the thin-shell thickness used in the spherical collapse computation, increasing the efficiency of the chameleon suppression.
We also compare our results to the phenomenological PPF fit of~\cite{li:11b}, providing a theoretical motivation for the functional form assumed therefor.
Note that the $N$-body results at the low-mass end are contaminated by the inclusion of subhalos, which are not identified and removed in the SO halo-finder employed.
}
\label{fig:massfunctionfitsandnu}
\end{figure*}

\subsection{Excursion set theory}  \label{sec:excursionmassfct}

Having determined the first-crossing distribution $\phi(S,\denv)$ in~\textsection\ref{sec:excursion} and its environmental average $\langle \phi(S,\denv) \rangle_{\rm env}$ in~\textsection\ref{sec:environment}, where $\phi(S,\denv) \, \rmd S$ describes the fraction of mass enclosed in halos of masses corresponding to $S(M)\in[S,S+\rmd S]$, the halo mass function can be computed from
\bq
 \frac{\rmd n(M)}{\rmd M} \rmd M = \frac{\brhom(\zform)}{M}  \langle \phi(S,\denv) \rangle_{\rm env} \left| \frac{\rmd S}{\rmd M} \right| \rmd M.
\eq
We show the relative difference of the halo mass function predicted by the excursion set approach for $f(R)$ gravity outlined in~\textsection\ref{sec:excursion} and \textsection\ref{sec:environment} with respect to $\Lambda$CDM in Fig.~\ref{fig:massfunctionSTenvalpha}.

\subsection{Sheth-Tormen halo mass function} \label{sec:STmassfunction}

Sheth \& Tormen~\cite{sheth:99} introduced a modification of the Press-Schechter expression for the first-crossing distribution as a function of the peak threshold $\nu\equiv\deltac/\sqrt{S}$,
\bq
 \nu \, \phi(\nu) = A \sqrt{\frac{2}{\pi} a \, \nu^2} \left[ 1 + \left(a \, \nu^2\right)^{-p} \right] e^{-a\,\nu^2/2}. \label{eq:st}
\eq
Here, $A$ is a normalization parameter, i.e., $\int \rmd \nu \, \phi(\nu) = 1$, and $a=0.707$ and $p=0.3$.
Eq.~(\ref{eq:st}) is designed such that the halo mass function
\bq
 \frac{\rmd n(M)}{\rmd M} \rmd M = \frac{\brhom}{M} \phi(\nu) \frac{\rmd \nu}{\rmd M} \rmd M
\eq
matches results from $N$-body simulations and the modification can be motivated by excursion set theory with a moving barrier such as caused through ellipsoidal collapse~\cite{sheth:99b, sheth:01}.

We compute the halo mass function defined by the first-crossing distribution in Eq.~(\ref{eq:st}) for the different $f(R)$ models assumed in the $N$-body simulations in~\textsection\ref{sec:simulations}.
$\deltac$ is determined through the spherical collapse model in~\textsection\ref{sec:sphcoll}, becoming dependent on mass and environment and entering Eq.~(\ref{eq:st}) through the peak threshold $\nu$.
We compare our predictions for $f(R)$ gravity to their counterpart from $\Lambda$CDM  in Fig.~\ref{fig:massfunctionSTenvalpha}, showing the relative enhancements of the halo mass function caused by the $f(R)$ modifications in different local environments $\denv$ and the environment-averaged case assuming the Eulerian distribution of $\denv$ given in Eq.~(\ref{eq:eulerian}).
In Fig.~\ref{fig:massfunctionfitsandnu}, we also show results from using $\alpha=1/2$ in Eq.~(\ref{eq:extraforceafudge}) to increase the efficiency of the chameleon suppression determined by the thin-shell expression Eq.~(\ref{eq:thinshell}).

\subsection{Nonlinear PPF formalism} \label{sec:PPF}

Li \& Hu~\cite{li:11b} introduce a nonlinear parametrized post-Friedmann (PPF) description to determine the halo mass function for chameleon $f(R)$ gravity.
They phenomenologically interpolate between the linearized and suppressed regimes by introducing a chameleon PPF transition in the variance as
\bq
 S_{\rm PPF}^{1/2}(M) = \frac{S_{f(R)}^{1/2}(M) + \left(M/M_{\rm th}\right)^{\mu} S_{\Lambda{\rm CDM}}^{1/2}(M)}{1+\left(M/M_{\rm th}\right)^{\mu}}, \label{eq:PPF}
\eq
where $M_{\rm th}$ and $\mu$ are calibrated to fit simulation results.
Assuming the same initial conditions for the $f(R)$ and $\Lambda$CDM models, the variance $S_{f(R)}^{1/2}$ is determined from Eq.~(\ref{eq:variance}) with the linear power spectrum
\bq
 P_{f(R)}(a,k) = \left( \frac{D_{f(R)}(a,k)}{D(a)} \right)^2 P(a,k),
\eq
where the linear growth functions are derived from solving Eqs.~(\ref{eq:lineargrowtheq}) and (\ref{eq:fRlineargrowtheq}).
The PPF peak threshold in~\cite{li:11b} is then given by
\bq
 \nu_{\rm PPF} \equiv \frac{\deltacLCDM}{S_{\rm PPF}^{1/2}(M)},
\eq
which they subsequently use in the Sheth-Tormen expression Eq.~(\ref{eq:st}) to approximate the halo mass function.
We simultaneously fit $(\bar{M}_{\rm th},\mu)$, where $\bar{M}_{\rm th} \left( 10^6 \absfR \right)^{3/2}~\Msunh \equiv M_{\rm th}$, to the enhancements in the halo mass function obtained from the $N$-body simulations for $\absfR=10^{-4}, \; 10^{-5}, \; 10^{-6}$ described in~\textsection\ref{sec:simulations}, finding $\bar{M}_{\rm th}\simeq2.172\times10^{12}$ and $\mu\simeq1.415$.
We show the PPF fits for the enhancements in the halo mass function and the corresponding peak thresholds in Fig.~\ref{fig:massfunctionfitsandnu}.

Based on our results for $\deltac$ from the spherical collapse model in~\textsection\ref{sec:sphcoll} applied to the Sheth-Tormen halo mass function in~\textsection\ref{sec:STmassfunction}, as an alternative determination of $\nu_{\rm PPF}$, we suggest generalizing and redefining the PPF peak threshold as
\bq
 \nu_{\rm PPF} \equiv \left\langle \frac{\deltac(M,\denv)}{S^{1/2}(M)} \right\rangle_{\rm env},
\eq
where $\langle\cdot\rangle_{\rm env}$ denotes the environmental average.
Note that $\deltac$ is determined using the linear $\Lambda$CDM growth function to extrapolate the initial overdensity associated with the collapse and $S^{1/2}$ is the variance obtained for $\Lambda$CDM.
In this definition, the chameleon transition is incorporated within $\delta_{\rm c}(M,\denv)$ through the estimation of the thin-shell thickness Eq.~(\ref{eq:thinshell}).
The advantage of this approach is that it is theoretically well motivated, that $\nu_{\rm PPF}$ may be determined without calibration of fitting parameters to simulation results, and hence,  that it encompasses dependencies on cosmological parameters and can easily be applied to other chameleon theories.
We compare the different approaches for computing $\nu_{\rm PPF}$ in Fig.~\ref{fig:massfunctionfitsandnu}, finding a good qualitative agreement between them and supporting the functional shape suggested in the phenomenological PPF interpolation formula Eq.~({\ref{eq:PPF}}).


\section{Conclusion} \label{sec:conclusions}

We have studied the spherical collapse of a top-hat overdensity in $f(R)$ gravity, taking into account the chameleon suppression of modifications in high-density regions.
The chameleon mechanism is approximated by an estimate of the thickness of a thin shell interpolating the scalaron field between the constant spherical halo overdensity and the constant spherical environmental density.
We implement this thickness estimation to approximate the nonlinear evolution of the spherical overdensity and the initial overdensity associated with the collapse.
The collapse density obtained by this procedure is environment- and mass-dependent.

We use excursion set theory to obtain the halo mass function predicted by $f(R)$ gravity and compare it to results from $N$-body simulations.
We further apply the peak threshold predicted by chameleon $f(R)$ gravity to the Sheth-Tormen fitting function for the halo mass functions of $\Lambda$CDM $N$-body simulations, to describe the enhancement of the $f(R)$ halo mass function relative to its $\Lambda$CDM counterpart.
Thereby, halo mass functions are predicted for different environments, where we assume an Eulerian environment distribution to estimate an averaged result.
Introducing a fudge factor in the thin-shell thickness to account for oversimplistic assumptions and approximations in the derivation of the chameleon barrier and to modulate the efficiency of the chameleon suppression, we can improve the description of the enhancement at the high-mass end of the halo mass function observed in the simulations.
This fudge factor is also preferred in the description of the difference between the lensing and dynamical mass of dark matter halos inferred from simulations~\cite{li:12}.

Finally, we compare our results to a nonlinear PPF fit, which introduces a description of the chameleon mechanism by interpolating the variance of the matter fluctuations between the linearized $f(R)$ gravity regime and the fully suppressed limit, corresponding to $\Lambda$CDM.
We find that the peak threshold predicted by our environment- and mass-dependent spherical collapse computations, averaged over the Eulerian environment, is in agreement with the peak threshold of the nonlinear PPF description, supporting the functional form suggested for this phenomenological fit.
While the PPF interpolation parameters have been fitted to $N$-body simulations using particular cosmological parameters, however, our derivation of the peak threshold in $f(R)$ gravity may be applied free of fitting parameters and it furthermore incorporates cosmological parameter dependencies.
Hence, our results can be used to extrapolate simulations beyond the set of simulated cosmological parameters for the use in parameter estimation analyses for inferring constraints on $f(R)$ gravity, employing observations sensitive to the cluster abundance.


\section*{Acknowledgments}

We thank Michael Kopp, Tsz Yan Lam, and Francesco Pace for useful discussions.
LL and KK are supported by the European Research Council,
BL by the Royal Astronomical Society and Durham University, and
GBZ by the Dennis Sciama Fellowship at the University of Portsmouth.
KK further acknowledges support from STFC (grant no.~ST/H002774/1 and ST/K0090X/1) and the Leverhulme trust.
$N$-body simulations have been conducted on computer facilities provided by the Western Canada Research Grid and the Sciama High Performance Compute cluster, which is supported by the ICG, SEPnet, and the University of Portsmouth.
Further numerical computations have been performed with ${\rm Maple}^{\rm \tiny TM}~16$ and Wolfram $Mathematica^{\rm \tiny \textregistered}~9$.
Please contact the authors for access to research materials.


\vfill
\bibliographystyle{arxiv_physrev}
\bibliography{fRmassfunction}

\end{document}